# Development and Characterization of a Microwave Atmospheric Pressure Plasma Jet

Suryasunil Rath and Satyananda Kar

*Plasma Applications Laboratory, Department of Energy Science and Engineering, Indian Institute of Technology Delhi, Hauz Khas, New Delhi, India-110016*

## Abstract

Over the past decade, atmospheric pressure discharges in the microwave frequency range have gained significant attention due to their promising applications in material processing, $CO_2$ dissociation, waste management, hydrogen production, water treatment, and more. This study presents the development and characterization of a waveguide-based microwave atmospheric pressure plasma jet (MW-APPJs), focusing on its design, diagnostics, and operational parameters. The setup incorporates a microwave power source, microwave waveguide networks, including the applicator section, and diagnostic tools for measuring plasma properties. Optical emission spectroscopy (OES) is employed to analyze the reactive species and determine plasma parameters which include electron excitation temperature ($T_{exc}$) and electron number density ($n_e$). The characterization highlights the influence of spatial and temporal gradients, gas flow rates, and power input on plasma behaviour. From OES, the $T_{exc}$ and $n_e$ variations were against the power increment. The thermocouple variations are also plotted with power.

## 1 Introduction

Atmospheric pressure plasmas provide several advantages over low-pressure plasmas, including the elimination of complex vacuum systems, simplified design, and ease of operation. Atmospheric Pressure Plasma Jets (APPJs) exhibit unique properties due to the generation of reactive oxygen and nitrogen species (RONS), such as OH, NO, $NO_2$, $ONOO^-$, $O_3$, $H_2O_2$, $HO_2$, $O_2^-$, and others. These RONS are primarily produced through interactions with ambient atmospheric gases and water vapor ($H_2O$) [1][2]. To maximize their effectiveness for various applications, plasma parameters must be carefully optimized within specific ranges. There are several methods for generating atmospheric pressure plasma discharges, including direct current (DC), alternating current (AC), radio frequency (RF), microwave (MW), and pulsed power sources [3][4][5][6]s. Each method offers distinct characteristics and different applications. To harness their full potential, a comprehensive understanding of their operational characteristics and physical properties is essential [7][8][9]. This requires a well-designed experimental setup and systematic diagnostic methods for effective characterization. The experimental setup of microwave atmospheric pressure plasma jets (MW-APPJs) involves careful integration of components such as microwave generators, plasma jet applicators, and diagnostic tools to ensure stable plasma generation and precise measurements [10][11]. Characterization of MW-APPJs focuses on assessing parameters such as temperature, electron density, and emission spectra, enabling insights into the complex physics governing these plasmas [12]. By understanding the experimental setup and characterization methods, MW-APPJs can be optimized for various advanced applications.

Section 2 details the design and configuration of the experimental set-up and plasma generation in MW-APPJs. Section 3 discusses surface wave production in microwave plasma. Section 4 focuses on the characterization of plasma jets by optical emission spectroscopy (OES) and thermocouple measurement. Section 5 discusses the results of the experiments, and finally, Section 6 presents the conclusion and outlines potential future work.

## 2 Experimental setup

Among APPJs, waveguide-based MW-APPJs are used for high-power handling capabilities. For our experiment, a standard rectangular waveguide (WR340) with (3.4 × 1.7) $inch^2$ or (8.636 × 4.318) $cm^2$ was taken, which ensures $TE_{10}$ mode propagations. The setup for MW-APPJs is shown in Figure 1. The magnetron connected to the power supply produces microwave (2.45 GHz) frequency by receiving the power from the power supply. The magnetron is connected to an isolator which protects the source from the backflow of the microwave. Next to the isolator, a triple waveguide (WR340) stub tuner is connected, which is used as a matching network. After the triple stub tuner, the applicator section is present in which the formation of plasma takes place. The applicator section consists of a tapered part that is used for the enhancement of electric field strength inside the waveguide. The waveguides are designed according to the guided wavelength ($\lambda_g$) which is the wavelength inside the rectangular waveguide. The expression of $\lambda_g$ is given by

$$\lambda_g = \frac{\lambda_0}{\sqrt{1-\left(\frac{\lambda_0}{2a}\right)^2}} \quad (1)$$

Where $\lambda_0$ is the wavelength of EM wave in free space, and $a$ is the width of the rectangular waveguide (WR340). For rectangular waveguide (WR340), the value of $\lambda_g$ is 17.42 cm. In the middle of the applicator, the waveguide is designed in such a way that the plasma breakdown will be at $\lambda_g/4$ distance. The choice of the $\lambda_g/4$ length can be explained through the principles of electromagnetic (EM) wave propagation in waveguides. At a distance of $\lambda_g/4$ within the waveguide, the wave amplitude reaches its maximum, corresponding to the highest amplitude in the wave profile. This high amplitude field strength facilitates more efficient plasma breakdown. The applicator section of the waveguide is designed to create a symmetric wave profile, ensuring that the maximum amplitude of the reflected EM wave (at $\lambda_g/4$) from the sliding short end is effectively coupled back into the plasma. The tapered section has a metallic enclosure (Faraday cage) that extends on both sides to prevent leakage of microwaves. In between the Faraday cage, a quartz tube is inserted in which plasma formation takes place by using the gas flow, as shown in Figure 1. For the experiment, a quartz tube with an inner diameter (ID) of 6 mm, an outer diameter (OD) of 10 mm, and a length of 400 mm was used. The quartz tube length of 400 mm can be axially adjustable inside the Faraday cage.

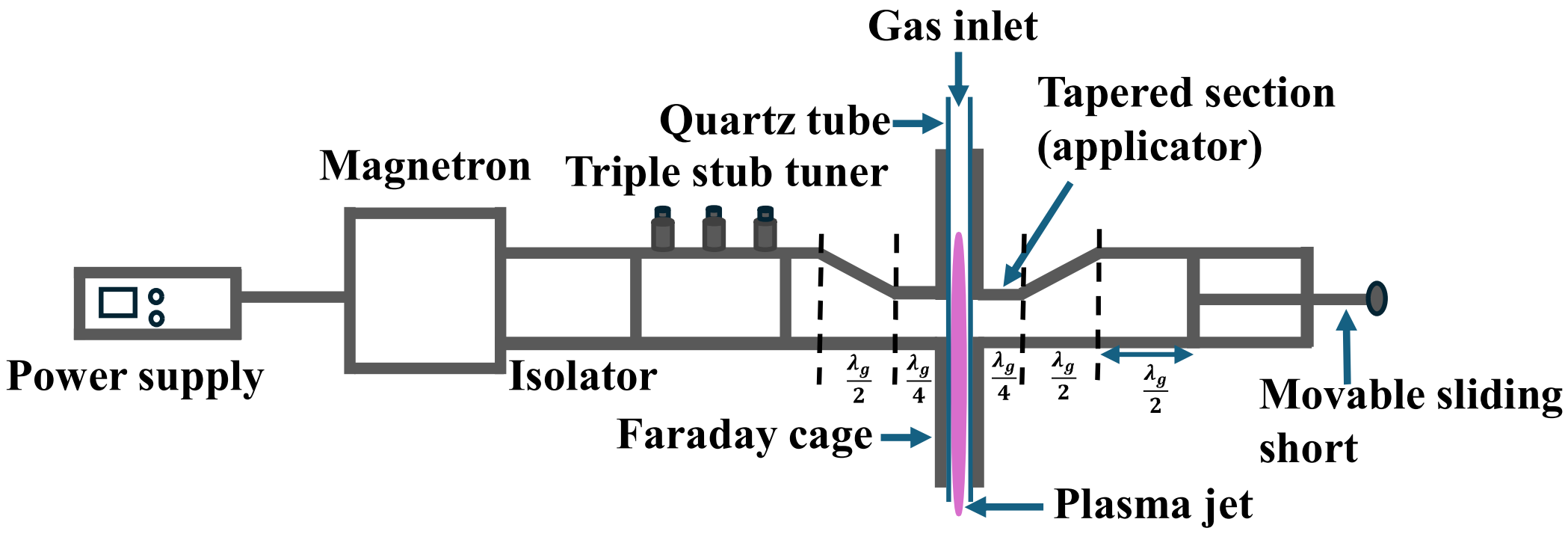

Figure 1. Schematic diagram of MW-APPJs

A sliding short is connected next to the applicator section for tuning purposes. Microwaves reflected from the sliding short couple back into the applicator section. For cooling, a water flow rate of 6 lpm is maintained for the power supply and magnetron, while a 2 lpm flow cools the stub tuners and applicator section. Additionally, a 15 lpm swirl airflow is used to cool the quartz tube in the applicator. Once all cooling systems are active, power is supplied to the waveguide to generate plasma. Typically, 100 W is required to ignite the plasma, with power increasing to 500 W for sustained plasma plume out from the applicator, which is enclosed in a Faraday cage. The quartz tube has an inner diameter of 6 mm and an outer diameter of 10 mm. The next section discusses surface wave generation in MW-APPJs.

## 3 Surface wave production in MW-APPJs

As the microwave-generated plasma has a higher frequency than the incoming microwave frequency, the wave suffers the skin effect, and the incoming microwave propagates as a surface wave. After the formation of plasma, due to the skin effect, incoming microwave (2.45 GHz) frequency cannot propagate inside, as followed by this equation:

$$\omega_p = \sqrt{\frac{n_e e^2}{m_e \epsilon_0}} \tag{2}$$

where, $\omega_p$ is the plasma frequency, $n_e$ is the electron number density, $e$ is the charge of an electron, $m_e$ the mass of electrons and $\epsilon_0$ is the permittivity of free space. From equation 2, $\omega_p$ increases with $\sqrt{n_e}$ i.e., high electron number density leads to higher plasma frequency. When the plasma frequency exceeds the applied frequency (2.45 GHz), the skin effect prevails, preventing wave penetration into the plasma. Instead, the waves propagate along the interface between the plasma and the dielectric medium as surface waves. Initially, the microwave coupling to a plasma created a space wave region around the applicator, later due to a guiding metallic section (Faraday cage) and skin effect, the space waves converted to surface waves [13]. The surface wave interacts with the lower-density plasma region, where the skin effect is negligible. The surface wave is a unique feature in MW-APPJs, which can increase the stability of plasma compared to other APPJs. The details of power coupling in microwave plasma can be found in [14][15][16].

## 4 Characterizations

MW-APPJs represent a distinct class of plasma that operates in a higher frequency range than other APPJs with the production of stable plasma due to surface waves. Characterizing plasmas at atmospheric pressure is notably more challenging than at low pressure due to the presence of pronounced temporal and spatial gradients. Unlike other plasma sources, MW-APPJs can generate highly reactive plasma without the need for vacuum environments, making it more practical for many real-world applications. For usefulness in different applications, characterization is very essential [17].

### 4.1 OES

OES is a widely used diagnostic method for MW-APPJs, valued for its simplicity and noninvasive nature [18]. The light emission from the plasma was collected via an optical fiber, which was connected to a spectrometer. The identification of spectral lines was primarily based on wavelength and relative emission intensity, referencing the National Institute of Standards and Technology (NIST) database. To facilitate spatially resolved measurements, a lens can be positioned at the tip of the optical fiber. In our experiments, electron excitation temperature ($T_{exc}$), and electron number density ($n_e$) by using line intensity ratio methods were calculated. For our OES spectra, Ocean Optics (HR4Pro) with 0.9 nm resolution was positioned, which covers the wavelength range 200-1100 nm. The OES data taken for power 600-1000 W, 3 lpm is shown in Figure 2.

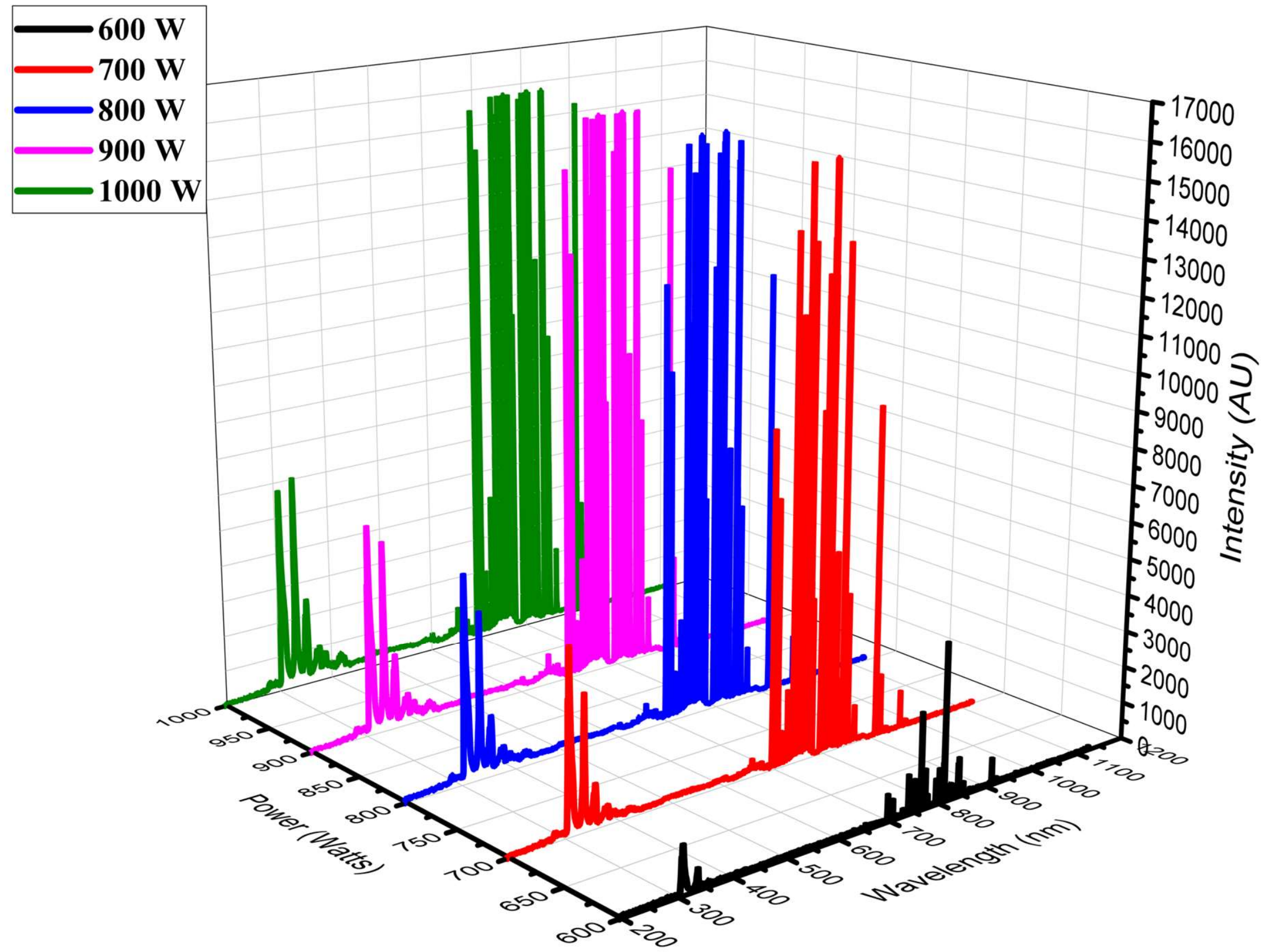

Figure 2. OES for 600-1000 W.

#### 4.1.1 Electron excitation temperature

The electron excitation temperature ($T_{exc}$) in plasma is a critical parameter used to describe the energy distribution of electrons in plasma and provides insights into its physical and chemical behaviour. It provides a window into the energy dynamics and reactivity of plasma, serving as a key parameter for diagnosing and controlling plasma behaviour across various applications. The calculation of the $T_{exc}$ assumption is based on the thermodynamic equilibrium of plasma given by the Boltzmann relations [19]. According to the Boltzmann relation,

$$ln\left[\frac{I_{ki}\lambda_{ki}}{\left(\frac{g_k}{g_i}\right)A_{ki}}\right] = ln(hcN_k) - \frac{1}{k_B T_{exc}}\,\Delta E_k \quad (2)$$

Where the subscript $ki$ means transitions from $k$ to $i$ level. $I_{ki}$ is the intensity and $\lambda_{ki}$ is the wavelength of transition. The $g_k$, and $g_i$ are statistical weights of upper and lower levels. $A_{ki}$ is the Einstein coefficient. The $h$ and $c$ are the Plank's constant and speed of light in vacuum, respectively. The $N_k$ is the species populations of the upper level, $k_B$ is the Boltzmann consant, and $E_k$ is the energy level of the upper states. $T_{exc}$ is given by the inverse slope of the plot of $ln\left[\frac{I_{ki}\lambda_{ki}}{\left(\frac{g_k}{g_i}\right)A_{ki}}\right]$ as a function of $E_k$. Figure 2 shows the OES data indicating different RONS and Ar lines. Based on Ar I transitions, the excitation temperature ($T_{exc}$) variations for 3 lpm are shown in Figure 3(a).

### 4.1.2 Electron number density ($n_e$)

The electron density, determined from the spectral lines of atoms and ions emitted by the plasma, is calculated using the Saha-Boltzmann equation [20][21], as given below

$$n_e = \frac{I_Z^*}{I_{Z+1}^*} 6.04 \times 10^{21} T^{\frac{3}{2}} exp\,[(-E_{k,Z+1} + E_{k,Z} - \chi_Z)/\,k_B T]\,cm^{-3} \tag{3}$$

$$I_Z^* = I_Z\, \lambda_{ki,Z}\,/\,g_{k,Z}\, A_{ki,Z} \tag{4}$$

Where the subscript $ki$ means transitions from $k$ to $i$ level. The $I_Z$ is the intensity of the transition, and $\lambda_{ki,Z}$ is the wavelength of transition in $Z$ state. The quantities $g_{k,Z,}$ and $A_{k,Z}$ are degeneracy of upper-level k in state $Z$ and transition probability, respectively. The $E_{k,Z}$ is the energy of transition in eV units. The quantity $T$ plasma temperature which is commonly interpreted as $T_{exc}$. The $\chi_Z$ is the ionization energy of species in state Z in eV units. Here the state Z corresponds to the excitation state of Ar atoms (Ar I, Z=0), whereas the Z+1 state is the ionized state of Ar (Ar II, Z=1). For our calculation of $n_e$, three Ar I is taken (727 nm, 826 nm, and 852 nm), whereas for Z+1 (Ar II) state is considered as 738 nm [22] . By averaging these three values, the variations of electron number density with power for 3 lpm are shown in Figure 3 (b).

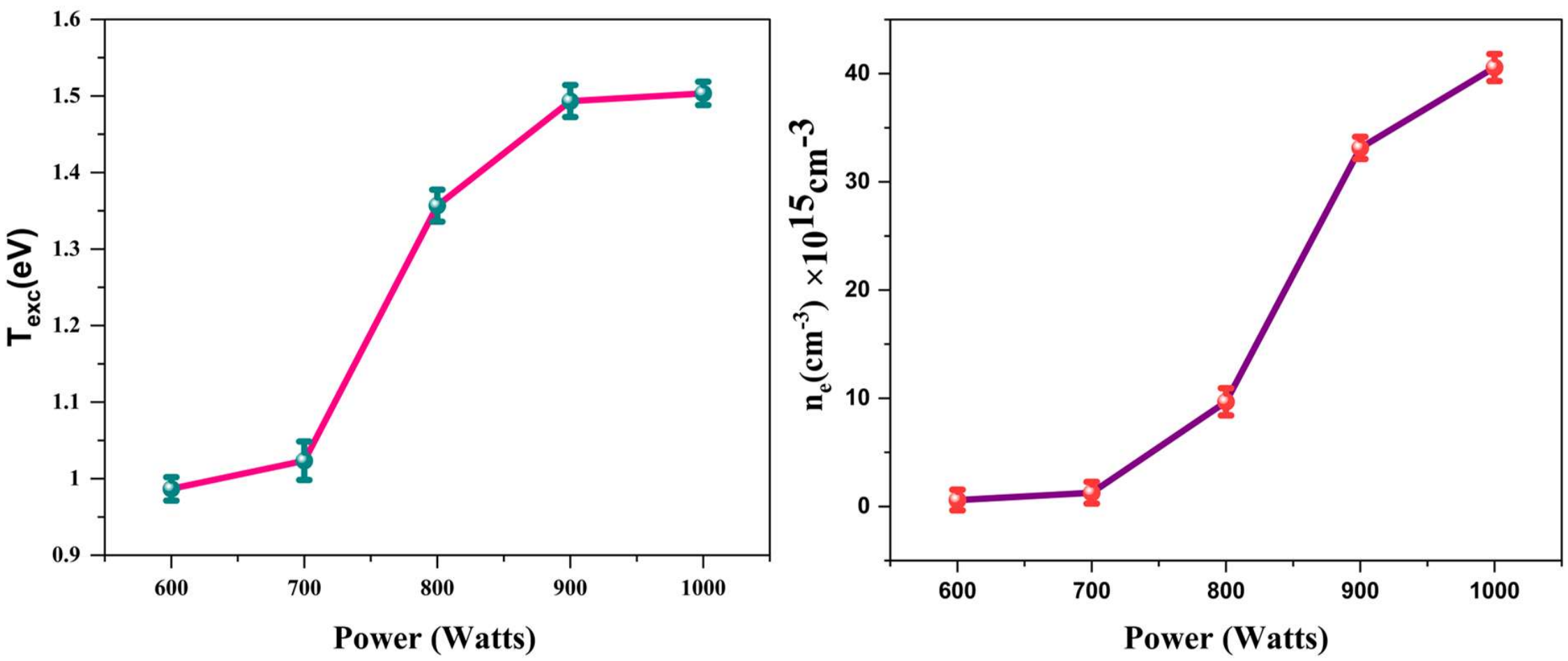

Figure 3. Variation with the power of a) electron excitation temperature, b) electron number density

### 4.2 Thermocouple measurement

Gas temperature measurements can be performed using a calibrated thermocouple or probe which is insulated from plasma, offering a simple and straightforward diagnostic method for atmospheric pressure plasmas. This technique operates on the principle of the Seebeck effect, which states that a temperature difference between two points in a conductive material can generate a corresponding voltage difference and vice versa. For our experimental system, an insulated K-type thermocouple was used which ranges from 0-1200$^0$ C. The thermocouple measurements for power 600-1000 W for different Ar gas flow rates are shown in Figure 4.

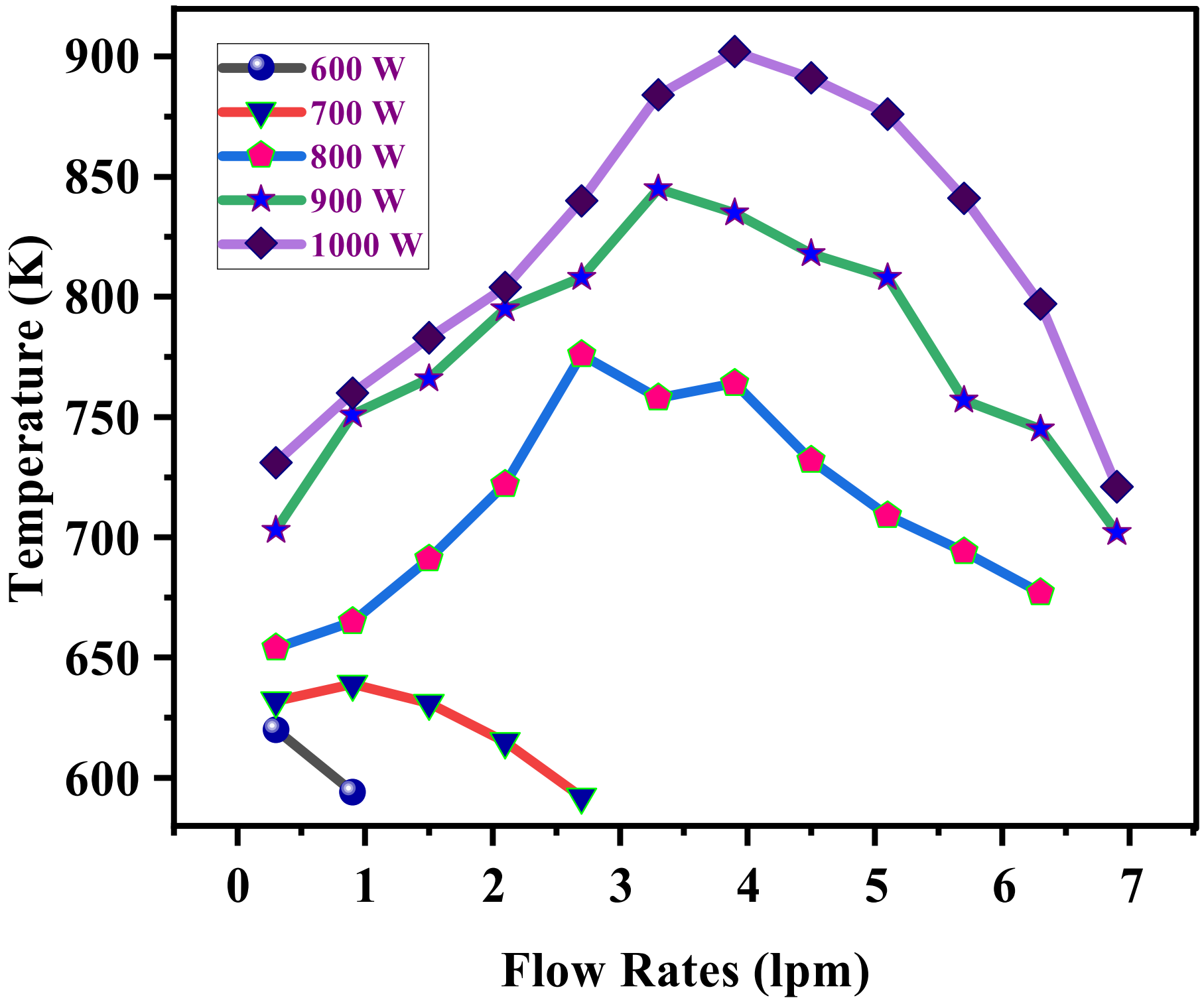

Figure 4. Variation of gas temperature with flow rates

## 5 Discussions

The plume length increases with increase with increase in power for fixed flow rates. This is due to a fix flow rate indicating the neutral Ar density is constant, then higher power will lead to higher ionizations. At constant power, the plume length was observed to initially increase with flow rates, but later it decreased for higher flow rates. This is due to a fix power, energy given to plasma is constant, but higher flow rates indicate high neutral Ar density, and for low flow rates, neutral Ar is much less, so power coupling will not be effective, whereas for higher flow rates indicate higher neutral density, power coupling also will be inefficient for both cases. So, for a fixed power, optimum ionizations depend on the flow rates of Ar gas.

From Figure 2, with an increase in power, the Ar line emission intensity increases due to higher ionizations of Ar. The RONS species is also increasing with increasing power. The increase in RONS species with power is due to a higher rate of energy transfer from the Ar atom to the atmosphere. The calculation of $T_{exc}$ variations were shown from 600-1000 W, 3 lpm. While increasing with power from 600 W to 1000 W, the $T_{exc}$ increases with the power, as shown in Figure 3 (a). This is due to the higher electron population in upper excited states after receiving higher energy from microwave power. After 1000 W, for the same positioning of optic fibre, the Ar I line saturates due to a higher rate of Ar I transitions. Similarly, $n_e$ also increases with power due to higher Ar atom ionization with higher power as shown in Figure 3 (b). The intensity of $H_\alpha$ (656.3 nm) increases from 600 W to 1000 W, as shown in Figure 2. The increase in $H_\alpha$ intensity indicates the higher interaction of plasma with atmospheric water vapor ($H_2O$) that is already present in the environment. Sometimes during experiments, to increase the intensity of $H_\alpha$ line, the Ar flow needs to pass with water vapor [23]. The findings of $H_\alpha$ are important for calculating the $n_e$ of atmospheric pressure plasmas. Thermocouple measurement directly gave the gas temperature ($T_g$). For a fixed power, temperature initially increases, attains a maximum value, and then decreases with higher flow rates as shown in Figure 4. At constant flow rates, a higher power results in an increased gas temperature ($T_g$). This is due to the optimum ionization rates for a given power, as discussed above.

## 6 Conclusion

In this study, the development and characterization of a waveguide-based MW-APPJs were successfully conducted. Key plasma parameters, including a higher electron excitation temperature, electron number density and elevated gas temperatures, were found. These findings highlight the capability of MW-APPJs to generate highly reactive environments suitable for various applications. The results demonstrate the effectiveness of the diagnostic methods and the potential of MW-APPJs for processes requiring high energy density and reactivity. Advanced diagnostic techniques with higher spatial resolution could provide deeper insights into plasma dynamics and gradients. Additionally, studies could focus on the interaction of MW-APPJs with various materials for applications in material treatment and environmental fields, such as targeted therapies and pollutant degradation. Numerical modelling could further enhance understanding and guide the optimization of plasma performance.

**Acknowledgements**

The author acknowledges the lab members of Plasma group IIT Delhi for their guidance and support.

---

Corresponding author

*E-mail: satyananda@dese.iitd.ac.in (Satyananda Kar)*